\documentclass[a4paper,11pt]{article}
\usepackage{pos}
\usepackage{hyperref}
\usepackage{xspace}
\usepackage{subcaption}  
\usepackage{CJKutf8}   

\newcommand{\pp}           {pp\xspace}

\newcommand{\PbPb}         {\mbox{Pb--Pb}\xspace}

\newcommand{\snn}          {\ensuremath{\sqrt{s_{\mathrm{NN}}}}\xspace}
\newcommand{\pt}           {\ensuremath{p_{\rm T}}\xspace}
\newcommand{\MeVc}         {Me\kern-.1emV/$c$\xspace}
\newcommand{\GeVc}         {Ge\kern-.1emV/$c$\xspace}
\newcommand{\TeV}          {Te\kern-.1emV\xspace}
\newcommand{\pTjet}        {\ensuremath{p_{\mathrm{T,\;jet}}}}
\newcommand{\pTchjet}      {\ensuremath{p_{\mathrm{T}}^{\mathrm{ch\; jet}}}}

\newcommand{\mjet}         {\ensuremath{m_\mathrm{jet}}}

\title{Measurement of the jet mass and angularities in \PbPb{} collisions at 5.02 \TeV{} with ALICE}
\ShortTitle{Jet mass and angularities in \PbPb{} collisions with ALICE}

\author*[a]{Ezra D. Lesser}

\affiliation[a]{University of California Berkeley\\
  Department of Physics\\
  Berkeley, CA, 94720-7300\\
  United States}

\emailAdd{elesser@berkeley.edu}

\abstract{
Jet substructure observables provide powerful tools to search for new
physics and test theoretical descriptions of perturbative and
non-perturbative processes in QCD. In heavy-ion collisions, jet
substructure observables are used to elucidate the structure
and dynamics of the quark-gluon plasma. Jet mass is one such observable, which
probes the virtuality of hard-scattered partons and their modified fragmentation.
Additionally, generalized jet angularities provide a powerful tool for
differential measurements of the jet shower and its modification, as
two parameters vary the weight of the jet constituents’ relative angle and \pt{}.
Previous measurements of the jet mass and jet angularities have 
shown conflicting differences in comparison with models.
To clarify these results, we present new measurements of the jet mass and jet angularities
using an identical jet sample. 
The high-precision tracking system of ALICE enables these measurements
over a broad range in \pt{}, with low-\pt{} reach that is unique at the LHC.
We report the generalized
jet mass and jet angularities using charged-particle tracks in \PbPb{}
collisions at $\snn = 5.02$ \TeV{}. Various jet angularity
parameters are investigated for the jet resolution parameter $R = 0.2$.
Results are compared to \pp{} collisions and theoretical models.
}

\FullConference{HardProbes2023\\
 26-31 March 2023\\
 Aschaffenburg, Germany\\}

\begin{document}
\maketitle

\section{Introduction}

Collisions of ultra-relativistic heavy ions at the Large Hadron Collider (LHC) allow study of collective effects in quantum chromodynamics (QCD) at high temperature and density. These collisions produce a strongly-interacting state of matter called the quark-gluon plasma (QGP)~\cite{Bjorken_QGP} in which quarks and gluons are deconfined. The hard scattering of two partons from these collisions forms collimated sprays of particles called jets, which traverse the QGP and are modified via interactions with the medium, an effect known as jet quenching~\cite{Bjorken_quenching, Appel_quenching, 1990_quenching, 1995_quenching}. Quenched jets can be used to probe the structure and development of the QGP, and provide information about QGP transport properties, degrees of freedom, and mechanisms for energy loss.

Jet substructure observables can directly quantify these QGP quenching effects. One such observable is the jet invariant mass, $\mjet \equiv \sqrt{E_\mathrm{jet}^2 - p_\mathrm{jet}^2}$, where $E_\mathrm{jet}$ and $p_\mathrm{jet}$ are the jet energy and momentum, respectively. The jet mass is a proxy for the virtuality $Q$ of the hard-scattered parton, which is larger for jets with broader fragmentation. The generalized jet angularities~\cite{Berger_2003_1, Berger_2003_2, Almeida_2009, Larkoski_2014} are another class of such observables, defined as
\begin{equation} \label{ang_eqn}
\lambda_\alpha^\kappa \equiv \sum\limits_{i \in \text{jet}}
\bigg( \frac{p_{\text{T},i}}{\pTjet} \bigg)^\kappa
\bigg( \frac{\Delta R_i}{R} \bigg)^\alpha
\equiv
\sum\limits_{i \in \text{jet}}
z_i^\kappa \theta_i^\alpha,
\end{equation}
where $i$ runs over constituents in the jet, $R$ is the jet resolution parameter, $p_\mathrm{T}$ is transverse momentum, and
$\Delta R_i \equiv \sqrt{(y_\text{jet} - y_i)^2 + (\phi_\text{jet} - \phi_i)^2}$
is the jet-constituent distance in the rapidity-azimuth ($y-\phi$) plane. The continuous, tunable parameters $\alpha$ and $\kappa$ define the specific angularity observable, which include the jet multiplicity $N_\mathrm{jet} = \lambda_0^0$, as well as the jet girth $g = \lambda_1^1 R$ and thrust $\lambda_{2}^{1}$, which are related by a varied angular weighting $\alpha$. These jet angularities are also theoretically related to \mjet{}~\cite{Kang_2018},
\begin{equation}\label{eq:thrust-mass}
\mathrm{jet\;thrust\;} \lambda_{2}^{1} =
\left( \frac{\mjet}{\pTjet R} \right)^2 + \mathcal{O}[(\lambda_2^1)^2].
\end{equation}

ALICE has measured both $g$ and \mjet{} in \PbPb{} collisions at $\snn = 2.76$ \TeV{} during LHC Run 1 and compared to \pp{} models~\cite{ALICE_mjet_2017, ALICE_ang_2018}. Significant quenching modification was observed for $g$, while no significant modification was seen for \mjet{}. The origin of this discrepancy is unclear: differing values of $R$ and \pTchjet{}, which vary the relative energy loss and nonperturbative dependence, versus varied values of $\alpha$, which change sensitivity to medium recoil effects and angular broadening, could both account for the difference. This conundrum has become known as the girth-mass puzzle.

More recently, ALICE has measured the infrared- and collinear-safe jet angularities (with $\kappa=1$ and $\alpha>0$)~\cite{Berger_2003} in \pp{} collisions at $\sqrt{s}=5.02$ \TeV{}~\cite{alice_jet_angularities}. Using these previous measurements as a baseline, these proceedings present new measurements of the angularities in \PbPb{} collisions at identical center-of-mass energy~\cite{ALICE_prelim} to systematically quantify in-medium substructure modifications. These angularities are compared with new measurements of \mjet{} in the same data, using equivalent jet definitions for the first time to address the girth-mass puzzle.

Section~\ref{sec:results} reports three conclusions from these physics measurements. Background-subtracted charged-particle jets are measured with $40 < \pTjet < 150$ \GeVc{}, extending the kinematic reach of previous measurements and probing the strength of jet-medium interactions at varying energy scales. Theoretical comparisons are given, which provide discrimination between models and inform future quenched jet substructure studies at the LHC.

\section{Experimental setup and analysis method}

The ALICE detector and its performance are described in~\cite{ALICE_experiment, ALICE_performance}.
These \pp{} data were collected using a minimum-bias trigger, requiring a coincidence in the two forward V0 scintillator detectors. The \PbPb{} data were collected using a high-multiplicity trigger to select 0-10\% centrality events~\cite{ALICE_centrality_HI}. The event selection includes a primary-vertex selection and the removal of beam-induced background events and pileup. Using both the ITS and TPC subdetectors, charged-particle tracks are reconstructed with $p_\mathrm{T} > 150$ \MeVc{} over pseudorapidity range $|\eta| < 0.9$.

Jets are reconstructed from charged-particle tracks with FastJet~\cite{fastjet} using the anti-$k_\mathrm{T}$ algorithm~\cite{antikt} and $E$-scheme recombination with resolution parameter $R = 0.2$. The $\pi^\pm$-meson mass is assumed for all jet constituents. In \PbPb{}, where jets have a large uncorrelated background, the event-by-event constituent subtraction method~\cite{background_sub} is used, with maximum recombination distance $R_\mathrm{max} = 0.1$. Results are unfolded via the iterative Bayesian algorithm~\cite{Bayes_unfolding} with a four-dimensional response matrix describing the detector and background response.

Dominant systematic uncertainties include uncertainty on the tracking efficiency and dependence on the model used for unfolding. These settings are varied, with the differences between the resulting distributions and nominal results taken as independent systematic uncertainties. Similar variations are performed for the unfolding regularization and binning, as well as the background subtraction parameters (for \PbPb{}). Background non-closure uncertainties are also evaluated for \PbPb{} data by unfolding simulated vacuum events embedded in a simulated thermal background.
\section{Results}
\label{sec:results}

\subsection{Necessity of a \pp{} baseline for jet quenching measurements}
\label{sec:pp_baseline}

\begin{figure}[t]
    \centering
    \includegraphics[width=0.414\textwidth]{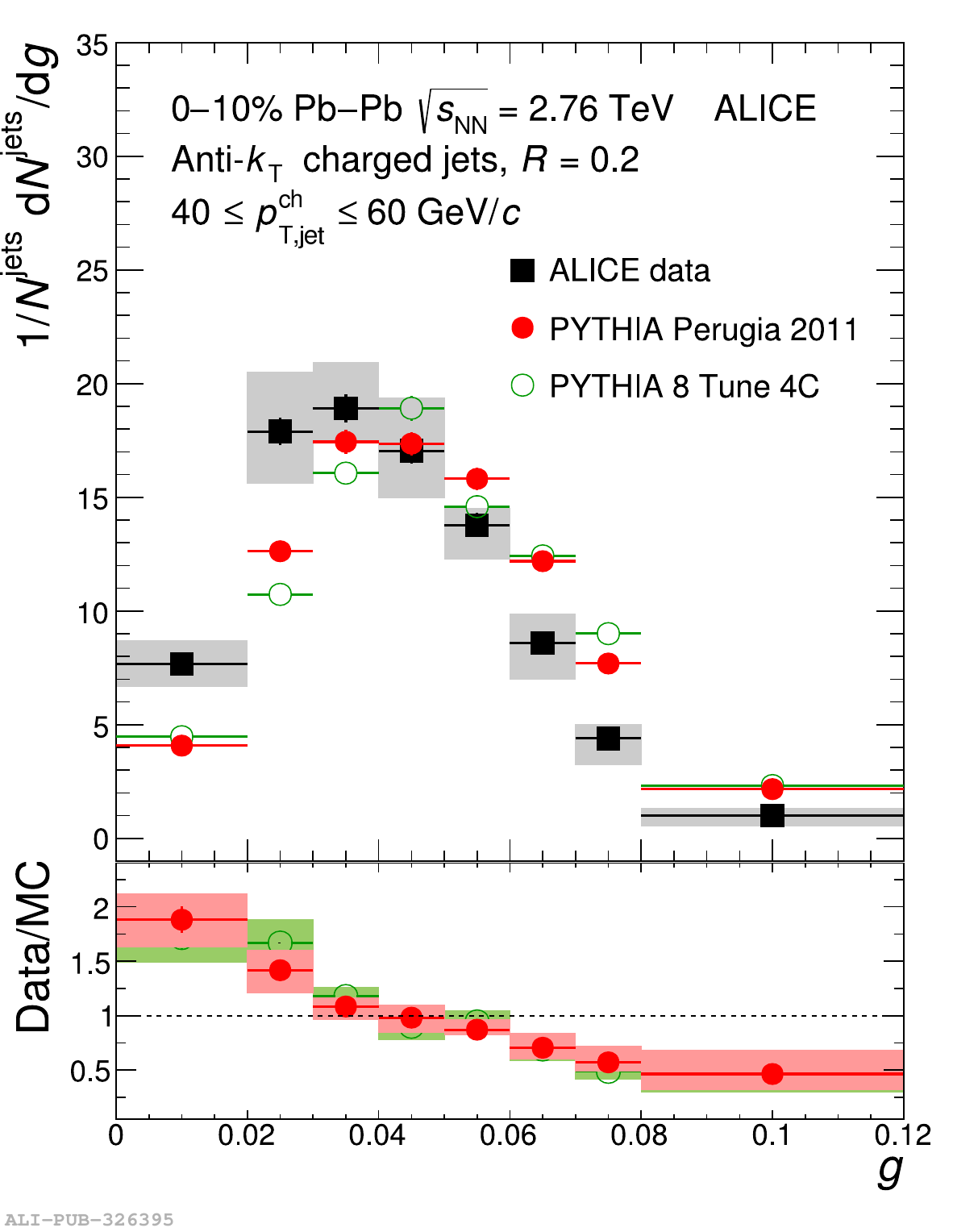}
    \hspace{2em}
    \includegraphics[width=0.3654\textwidth]{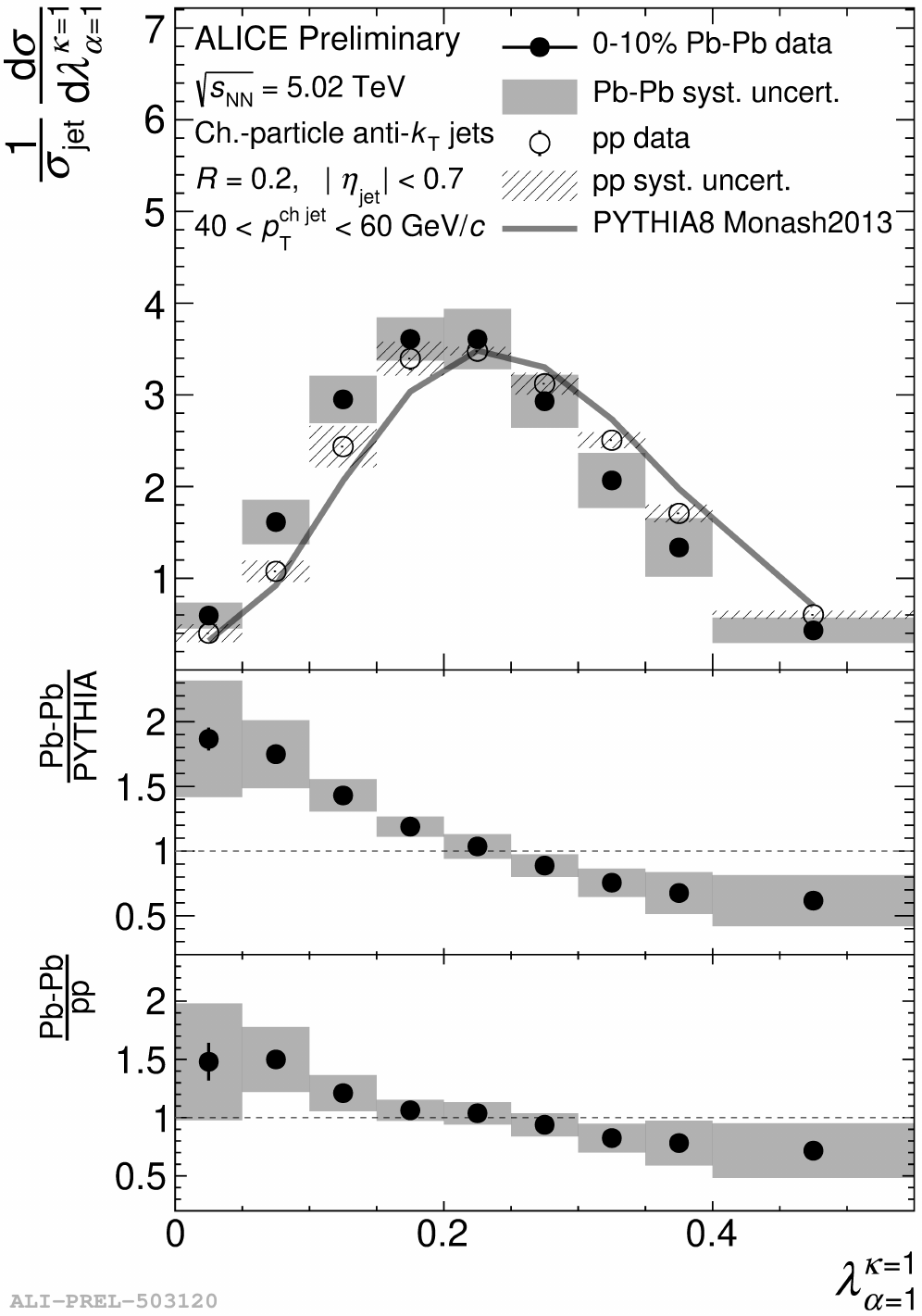}
    \caption{Measurement of the jet girth $g = \lambda_1^1 R$ in \PbPb{} data at $\snn = 2.76$ \TeV{} from LHC Run 1 (left) and at $\snn = 5.02$ \TeV{} from Run 2 (right). On the left figure, the bottom ratio panel compares data to PYTHIA baselines, as does the central ratio panel on the right figure. However, the comparison to a proper \pp{} baseline in the bottom ratio panel significantly pushes the ratio closer to unity.}
    \label{fig:girth_comp}
\end{figure}

ALICE has performed new measurements of $\lambda_1^1 = g / R$ using \PbPb{} data from LHC Run 2. Figure~\ref{fig:girth_comp} compares these results with those of $g$ from LHC Run 1~\cite{ALICE_ang_2018}. Comparisons of the data to baseline simulations from PYTHIA~\cite{PYTHIA} reveal strong modification in both datasets, with quenched jets exhibiting a ``narrowing'' behavior via enhancement at small values (or a corresponding suppression at large values) of angularity (girth), with both tails modified by an approximate factor of 2. However, the new result is also compared to a baseline of \pp{} data taken at equivalent center-of-mass energy, significantly reducing the narrowing. This baseline data was not available for the earlier result. The model-skewed ratio enhances perceived quenching of this observable without any underlying physical explanation. A proper \pp{} baseline is therefore essential for properly interpreting measurements of jet quenching in an unbiased way. This result has far-reaching implications for future runs at the LHC: heavy-ion data must pair with statistically consistent jet samples in \pp{}, where smaller collision systems result in fewer jets.

\subsection{Resolving the girth-mass puzzle}
\label{sec:girth-mass}

\begin{figure}[t]
    \centering
    \includegraphics[width=0.34\textwidth]{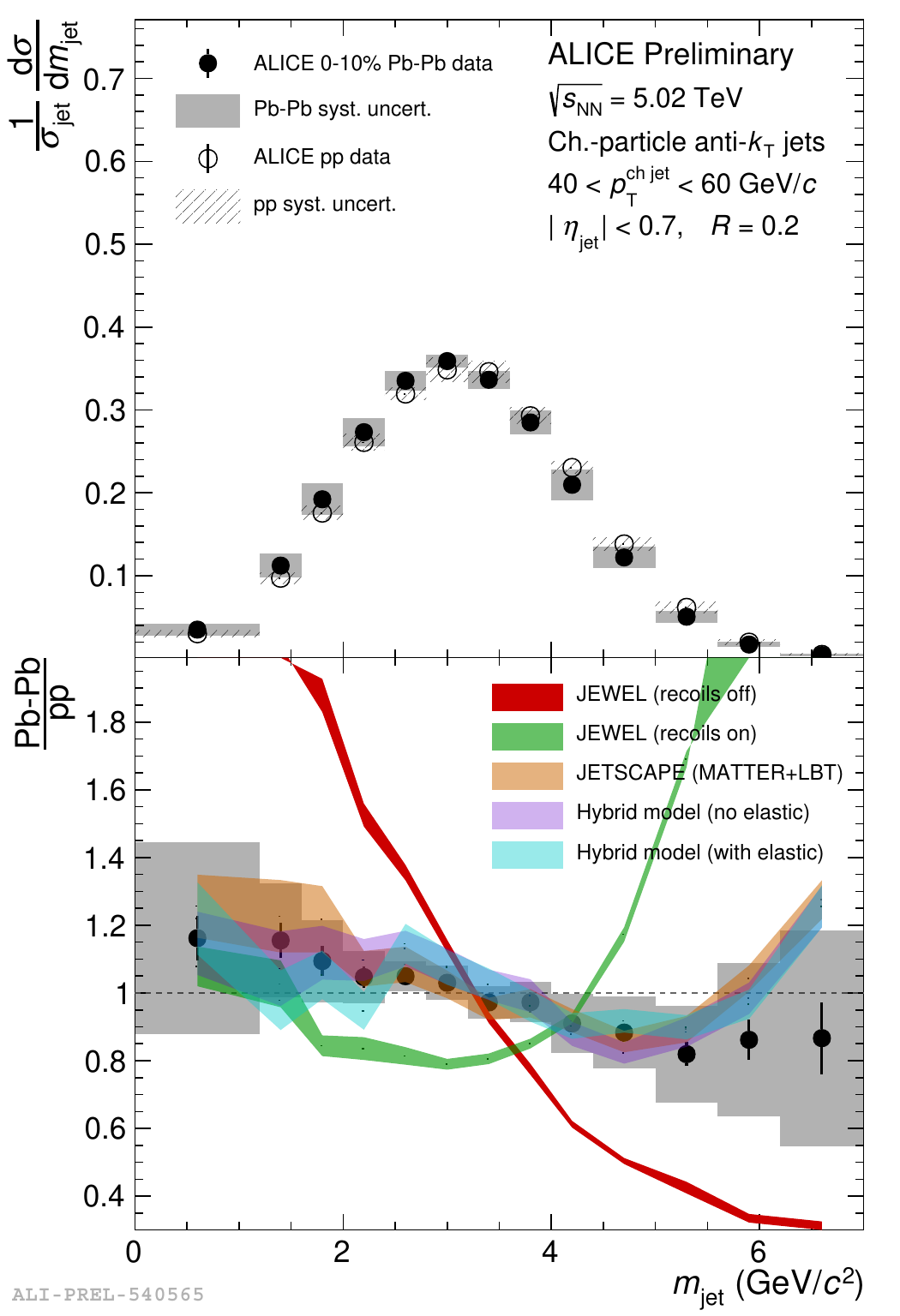}
    \hspace{2em}
    \includegraphics[width=0.34\textwidth]{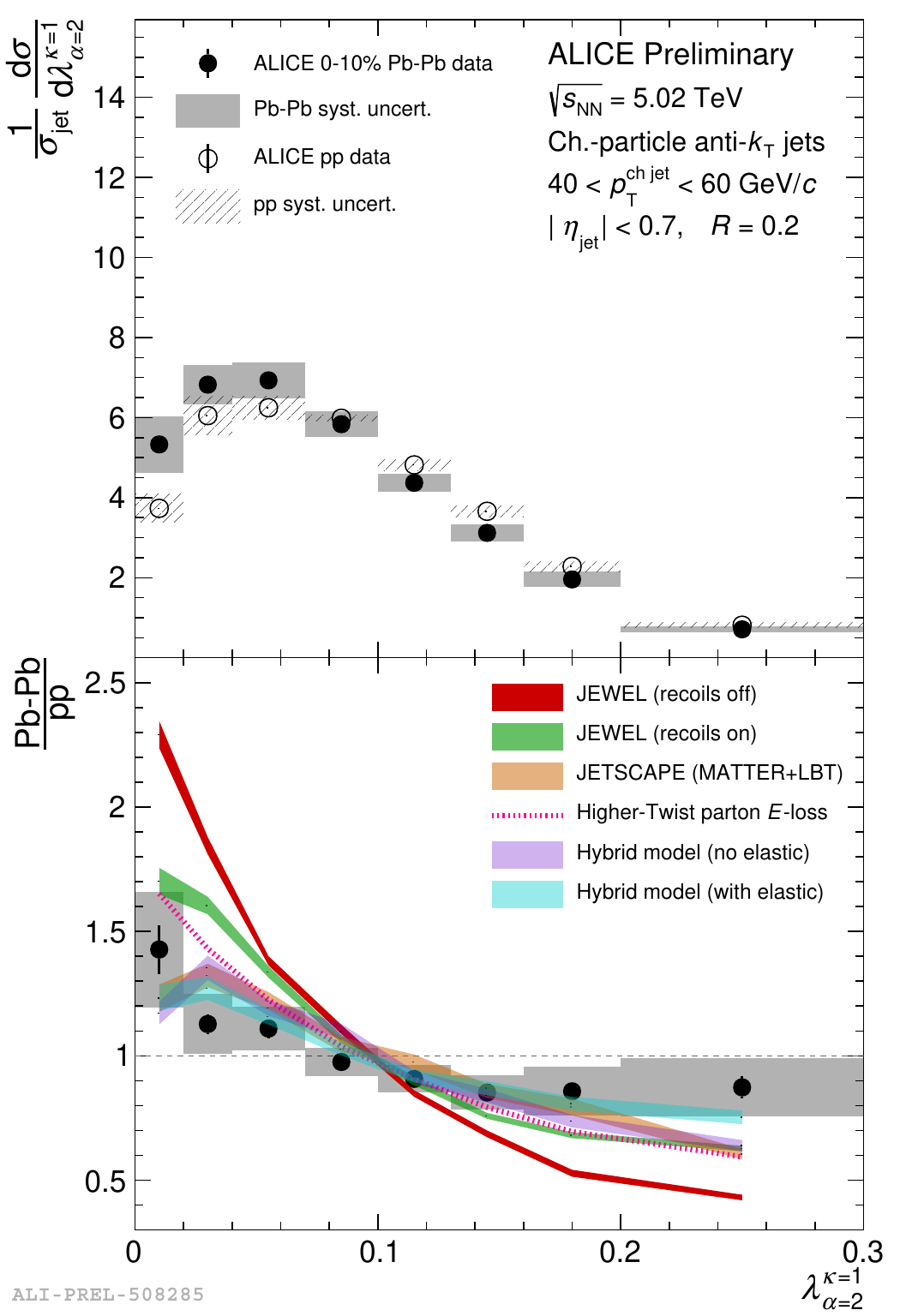}
    \caption{Measurement of the jet mass $\mjet$ (left) and thrust $\lambda_2^1$ (right) in \PbPb{} and \pp{} data at $\snn = 5.02$ \TeV{}. Systematic uncertainties are assumed to be totally uncorrelated between \pp{} and \PbPb{}. Comparisons are made to several models~\cite{JEWEL, Higher-Twist, jetscape, MATTER, LBT, Hybrid, Moliere}, which show varying behavior with respect to measured data.}
    \label{fig:thrust-mass}
\end{figure}

In order to study the girth-mass puzzle in light of Eq.~\ref{eq:thrust-mass}, ALICE has performed new measurements of \mjet{} and $\lambda_2^1$ using the same jet sample for the first time. Figure~\ref{fig:thrust-mass} compares the two distributions using identical \pTchjet{} intervals. While Eq.~\ref{eq:thrust-mass} relates \mjet{} and $\lambda_2$ directly to one another, model comparisons show differing behavior. JEWEL~\cite{JEWEL} (with recoils on), for example, overestimates enhancement at large values of \mjet{}, while it underestimates the yield at large jet thrust. Since the distributions are positive definite and obey square proportionality following Eq.~\ref{eq:thrust-mass}, large corrections to Eq.~\ref{eq:thrust-mass} must apply at these values of \pTchjet{}. This could include nonperturbative effects such as hadronization as well as higher-order correction terms $\mathcal{O}[(\lambda_2)^2]$. Despite their mathematical similarity, underlying physical differences between the two observables exist: the jet mass is sensitive to quark masses, whereas the IRC-safe jet angularities are sensitive to fragmentation and quark- versus gluon-initiated jet differences. Identifying the variations in the measured distributions as these physical differences of the observables explains the girth-mass difference.

This observation highlights the importance of making broad measurements of quenched jet substructure, as closely-related observables can provide significantly different probes of underlying physical phenomena. Studies of quenched jets using $N$-subjettiness variables as a basis suggest that dozens of such observables may be required to optimally characterize quenched jet behavior~\cite{info_content}.

\subsection{Grooming quenched jet substructure}
\label{sec:grooming}

\begin{figure}[t]
\center
\begin{subfigure}[b]{0.32175\textwidth}
    \centering
    \includegraphics[width=\textwidth]{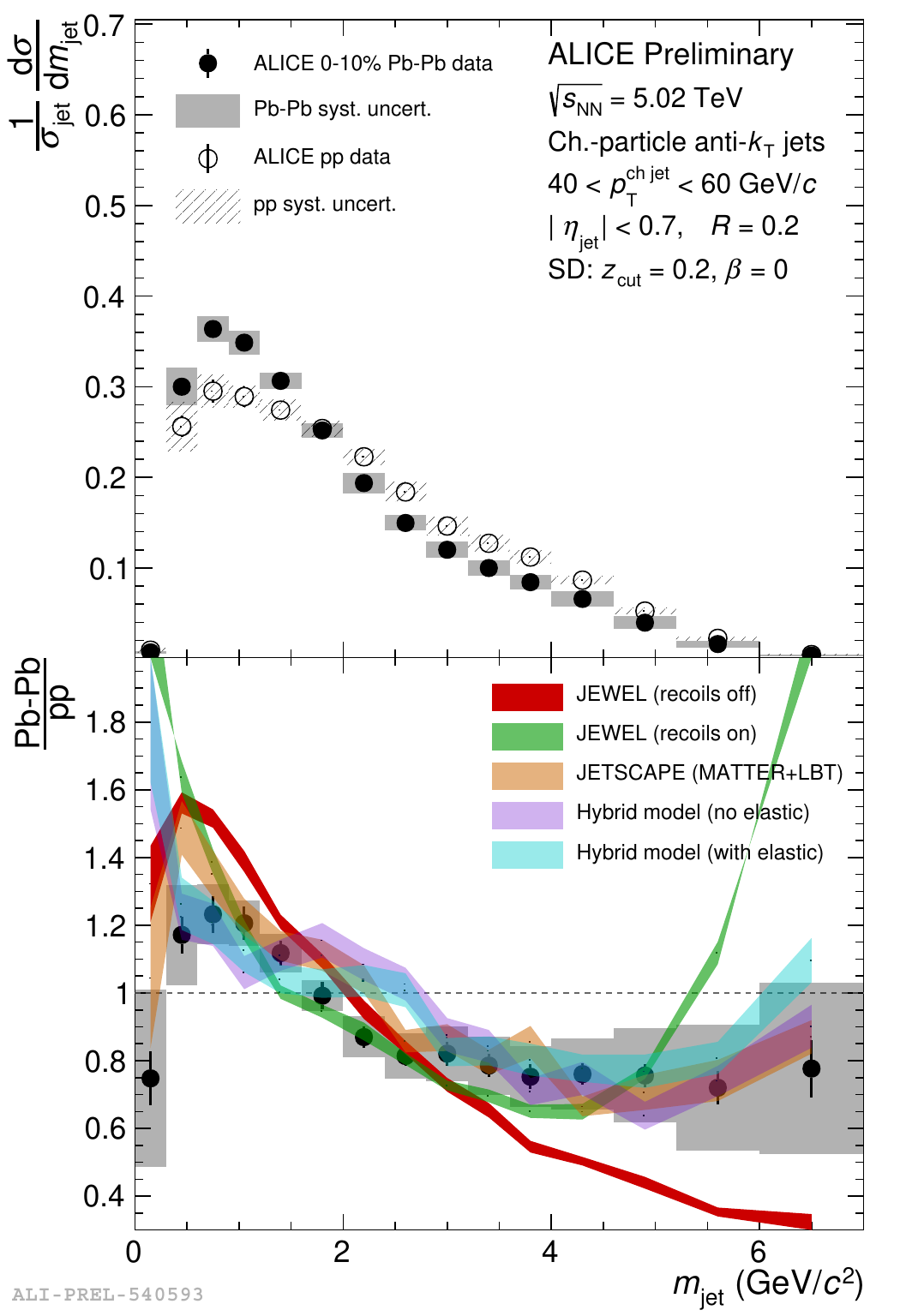}
    \caption{Groomed \mjet{} distribution.}
    \label{fig:ungr_vs_gr:mass}
\end{subfigure}
\hfill
\begin{subfigure}[b]{0.65\textwidth}
    \centering
    \includegraphics[width=0.495\textwidth]{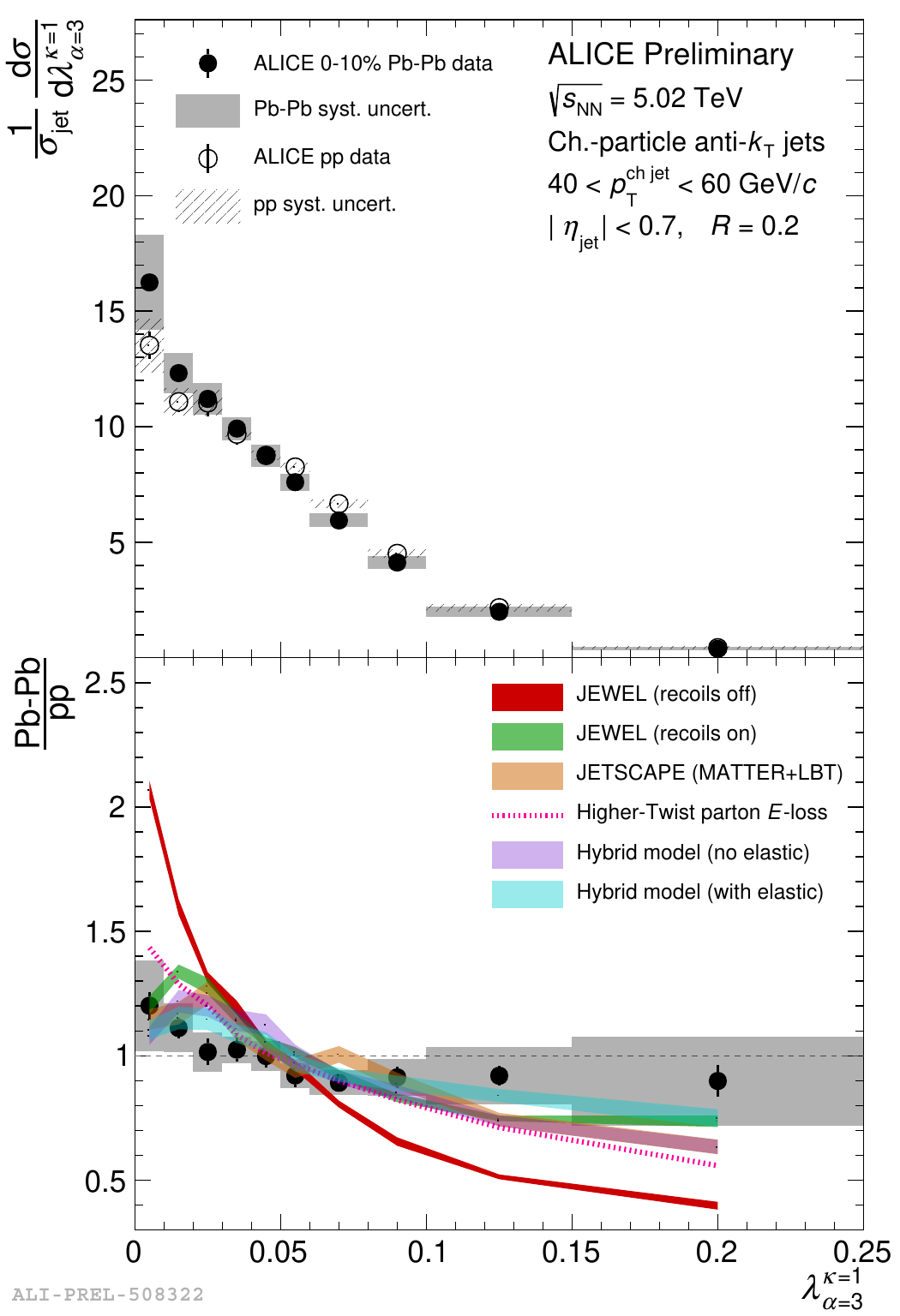}
    \includegraphics[width=0.495\textwidth]{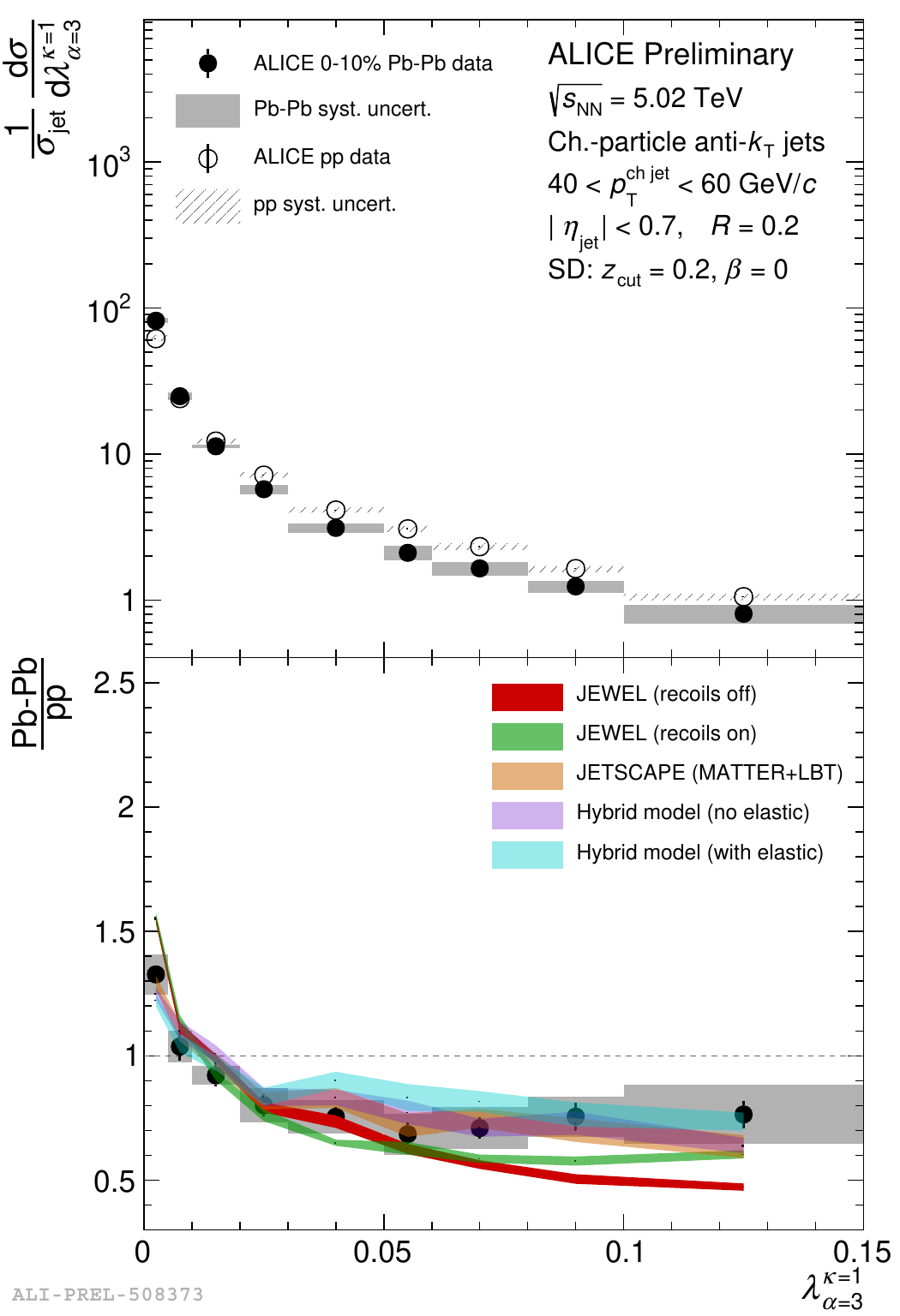}
    \caption{Ungroomed (left) vs. groomed (right) jet angularity $\lambda_{3}^1$.}
    \label{fig:ungr_vs_gr:girth}
\end{subfigure}
\caption{Jet angularity $\lambda_{3}^1$ and groomed \mjet{} results at $\snn = 5.02$ \TeV{} in $ 40 < \pTchjet < 60$ \GeVc{}.}
\label{fig:groomed}
\end{figure}

Jet grooming enhances the perturbative calculability of jet substructure observables, while its use in heavy-ion collisions additionally reduces contamination from the thermal background. Figure~\ref{fig:groomed} shows new groomed \mjet{} and $\lambda_3^1$ measurements from ALICE. Compared to their ungroomed counterparts, the groomed distributions display reduced systematic uncertainties and an enhanced narrowing effect, corresponding with a strongly quenched jet core. For the jet angularities, several model predictions converge with grooming, limiting differentiation between them despite incompatible theoretical approaches~\cite{JEWEL, Higher-Twist, jetscape, MATTER, LBT, Hybrid, Moliere}. Contrary to \pp{} measurements, where nonperturbative effects are optimally minimized to enhance tests of perturbative QCD, scrutinizing jet quenching models requires consideration of the significant nonperturbative effects in addition to the perturbative ones. 

\section{Conclusions}
\label{sec:conclusions}

ALICE has performed new measurements of the jet angularities and mass in \PbPb{} collisions at $\snn = 5.02$ \TeV{}. These results are compared to \pp{} data, unavailable for earlier LHC measurements. Quenching modification is reduced as compared to a simulated baseline, signaling that \pp{} data at comparable center-of-mass energy is essential for future LHC heavy-ion runs. Despite theoretical similarity, \mjet{} and the jet thrust $\lambda_2^1$ have differing behavior as compared to models, signifying large corrections to the leading-order perturbative equality and resolving the girth-mass puzzle. Grooming jets quenched by the QGP reveals enhanced narrowing of the hard jet core and reduced systematical uncertainties, while also reducing nonperturbative effects such as the thermal background. Increased similarity between models also suggests perturbative agreement and a continuing need for probing nonperturbative effects.

\newenvironment{acknowledgement}{\relax}{\relax}
\begin{acknowledgement}
\section*{Acknowledgements}
Many thanks to
Daniel Pablos Alfonso, Krishna Rajagopal,
James Mulligan, Yasuki Tachibana, Abhijit Majumder,
Shi-Yong Chen (\begin{CJK*}{UTF8}{gkai}陈时勇\end{CJK*}),
and Ben-Wei Zhang (\begin{CJK*}{UTF8}{gkai}张本威\end{CJK*})
for generating many theoretical predictions for these ALICE results and for engaging in useful discussions.
\end{acknowledgement}

\end{document}